\documentstyle[aps,prd,epsfig]{revtex}
\newcommand{\bee}{\begin{equation}}
\newcommand{\ee}{\end{equation}}
\newcommand{\beea}{\begin{eqnarray}}
\newcommand{\eea}{\end{eqnarray}}

\newcommand{\Tr}{\mbox{Tr}}

\begin{document}
\draft

\title{Witten-Veneziano Relation, Quenched QCD, and Overlap Fermions}
\author{Thomas DeGrand}
\address{
Department of Physics,
University of Colorado, 
        Boulder, CO 80309 USA
}
\author{Urs M. Heller}
\address{
CSIT, Florida State University, Tallahassee FL 32306-4120 USA
}
\author{(MILC Collaboration)}
\date{\today}
\maketitle
\begin{abstract}
The quarks in 
quenched QCD have an anomalous self-interaction in the flavor singlet Goldstone
boson channel. This coupling is extracted from a graph with disconnected
quark lines, and is used to infer the mass of the eta-prime meson
in full QCD. When the fermions are described by an overlap action,
the Witten-Veneziano relation is an exact relation
between the topological susceptibility
(as defined through fermionic zero modes)
and the inferred value of the eta-prime mass.
Using an overlap action
we compute the hairpin amplitude
and determine the fermion zero-mode susceptibility, the
inferred eta-prime mass and other parameters
characterizing the low energy chiral properties of quenched QCD.
\end{abstract}

COLO-HEP-479, FSU-CSIT-02-07

%

\section{Introduction}

The physics of the flavor singlet pseudoscalar eta-prime meson has
long been a source of puzzles.

\begin{figure}[thb]
\begin{center}
\epsfxsize=0.8 \hsize
\epsffile{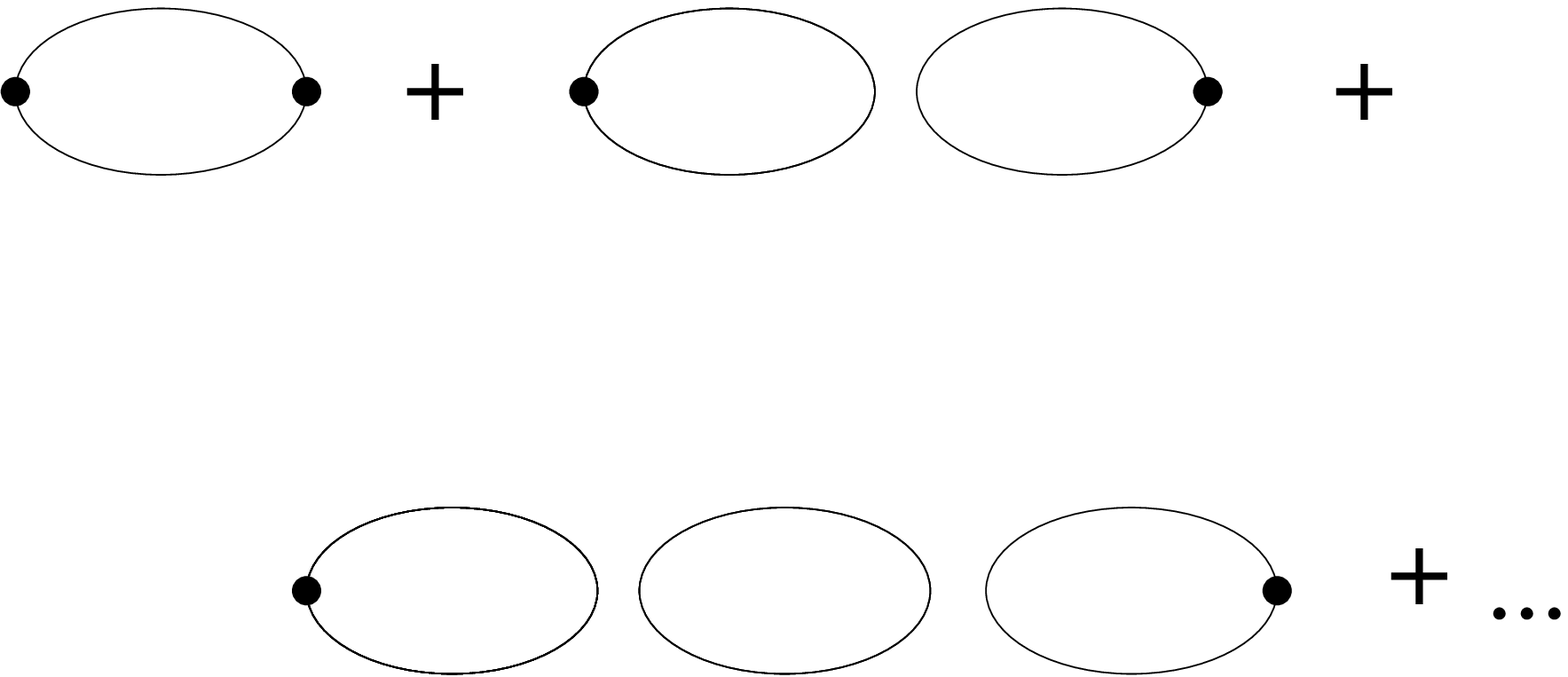}
\end{center}
\caption{
A plausible set of quark line graphs, which sum into a geometric series
to shift the eta-prime mass away from the mass of the flavor nonsinglet
pseudoscalar mesons. In the quenched approximation, only the first two
terms in the series survive as the ``direct'' and ``hairpin'' graphs.
}
\label{fig:graphs}
\end{figure}

\begin{itemize}
\item{ The eta-prime is not a Goldstone boson because of the presence of the
axial anomaly.
}
\item{The eta-prime propagator in full QCD involves a set of connected
 and disconnected quark-line diagrams. Summing a plausible subset of these 
diagrams (see Fig. \ref{fig:graphs})
 produces a shift in the eta-prime mass away from the masses
of the true Goldstone bosons. But what gauge field dynamics ``fills in 
the white space'' between the quark loops?
}
\item{
In the limit of a large number of colors $N_c$, the eta-prime mass is
 expected to scale  as $1/N_c$ as opposed to the masses of
 non-Goldstone mesons, which remain order one. A mass formula relating
the eta-prime mass to the topological susceptibility has been derived
by Witten and Veneziano
\cite{ref:WV}
\bee
m_{\eta'}^2 + m_\eta^2 - 2m_K^2 = \mu_0^2 = {{4 N_f \chi_T} \over {f_\pi^2}}
\label{eqn:WV}
\ee
where $\chi_T$ is the topological susceptibility of the pure gauge theory,
$N_f$ is the number of flavors,
and
the pion decay constant, defined through
 $\langle 0 | \bar \psi \gamma_0 \gamma_5 \psi |\pi\rangle = m_\pi f_\pi$,
has the experimental value of 132 MeV. The right-hand side is formally
 $O(1/N_c)$ because $f_\pi \simeq \sqrt{N_c}$. The suppression of 
fermion loops at large $N_c$ is similar to the absence of fermion loops
 in the quenched approximation, and lattice calculations of the topological
susceptibility in quenched QCD
(using pure gauge observables) give numbers with good
numerical agreement with Eq. (\ref{eqn:WV}), when evaluated with the physical
masses of the particles. It seems strange that this
large-$N_c$ formula should be quantitatively correct, since the eta-prime mass
is not particularly small compared to the masses of other non-Goldstone
boson mesons.
(It is worth remarking that the flavor-nonsinglet
pseudoscalar sector is also the home of other large-mass quantities:
one example is the scale of chiral symmetry breaking:
 $m_\pi^2/(m_u+m_d) \simeq 3$ GeV. The physical eta-prime mass is less
 than one third of that number.)
}
\end{itemize}
The relation of these statements to each other is an ongoing  source of
 controversy. Lattice simulations can in principle contribute to
the discussion and indeed, there have been many studies of the
eta-prime on the lattice, dating back to the earliest days of 
simulation (a partial list includes
 \cite{Hamber:1983vu,Fukugita:1984yw,Itoh:1987iy,Kuramashi:1994aj,Fukugita:1995iw,Venkataraman:1997xi,Bardeen:2000cz}).
Most of these studies have been in the quenched approximation, where
the mass of the eta-prime is inferred from the size of the hairpin graph
compared to the connected graph, and compared to the topological
 susceptibility. When the contribution of individual eigenmodes
of the Dirac operator to the hairpin graph is computed,
 it is often seen that the low-lying eigenmodes
make a substantial contribution to the hairpin graph.

This paper is another calculation of the hairpin diagram in quenched QCD.
Why revisit this question yet again? The reason is that the calculations
presented here are done with a fermion action which respects chiral
symmetry on the lattice via the Ginsparg-Wilson relation \cite{ref:GW}:
an overlap \cite{ref:neuberfer} action.  All previous studies of the
 eta-prime were
done with lattice actions which have chiral symmetry breaking artifacts:
for Wilson-type fermions, for example,
real eigenmodes of the Dirac operator are not zero modes nor are they
eigenstates of $\gamma_5$.
Since the properties of the eta-prime are known to be intertwined with
the anomaly, one would expect that calculations with exact lattice
chiral symmetry might be more revealing for the study of the eta-prime.

Indeed that is the case: when the fermions are described by an overlap action,
the Witten-Veneziano relation is an exact relation between the expectation
value of the hairpin diagram--which is given by the susceptibility of
fermion zero modes--and the
 anomalous coupling between two flavor singlet Goldstone bosons. Interpreting
that coupling as the eta-prime mass and equating the zero mode susceptibility
to the topological susceptibility yields Eq. (\ref{eqn:WV}).

To see this, first recall some facts about the overlap action:

 The eigenmodes of any massless
overlap operator $D(0)$
are located on a circle in the complex plane of radius $x_0$
with a center at the point $(x_0,0)$. The corresponding eigenfunctions are
either chiral (for the eigenmodes with real eigenvalues
located  at $\lambda=0$ or $\lambda=2x_0$)
or nonchiral  and paired; the two eigenvalues of the nonchiral modes
are complex conjugates.
 The massive overlap Dirac operator is
conventionally defined to be 
\bee
D(m_q) = ({1-{m_q \over{2x_0}}})D(0) + m_q
\ee
and it is also conventional to define the propagator so that the chiral
modes at $\lambda=2x_0$ are projected out,
\bee
\hat D^{-1}(m_q) = {1 \over {1-m_q/(2x_0)}}(D^{-1}(m_q) - {1\over {2x_0}}) .
\label{SUBPROP}
\ee
For a summary of  useful formulas, see Ref. \cite{ref:FSU98}
(for the special case $x_0=1/2$).

Now consider the hairpin diagram involving a single flavor of quarks,
 where the source and sink (black dots
in Fig. \ref{fig:graphs}) are the local pseudoscalar density,
 $\bar \psi \gamma_5 \psi$.
The hairpin diagram then is just
\bee
H(x,y) = \langle \Tr \gamma_5 \hat D(x,x)^{-1}
\Tr \gamma_5 \hat D(y,y)^{-1} \rangle .
\ee
Because only zero modes are chiral, the volume integral of the
hairpin graph is proportional to the zero mode susceptibility
\cite{ref:FSU98}
\bee 
{1 \over V} \sum_{x,y} H(x,y) =  {{ \langle Q^2 \rangle} \over {V m_q^2 }} =
 { \chi \over m_q^2 } ,
\label{eqn:HAIR1}
\ee
where $Q$ is just the number of zero modes,
 the difference of positive and negative chirality zero modes, $Q=n_+ - n_-$.
Regardless of any dynamical model used to describe the hairpin graph,
we expect to see a large contribution from zero modes
to $H(x,y)$ itself, since only they contribute
to the susceptibility.

In quenched QCD, as described by 
quenched chiral perturbation theory \cite{ref:chiralpt},
 there is  an anomalous coupling of two Goldstone bosons
 in the flavor singlet channel,
parameterized by a coupling with the dimensions of a squared mass.
The hairpin graph is analyzed as if each of its quark
loops is a propagator for an ordinary pseudoscalar Goldstone meson.
That is, the momentum space amplitude for the
 connected graph is
\bee
C(q)= f_P {1 \over {q^2 + m_\pi^2}} f_P
\ee
while the hairpin amplitude involving a single flavor is
\bee
H(q) = f_P {1 \over {q^2 + m_\pi^2}}
 {\mu_0^2 \over N_f} {1 \over {q^2 + m_\pi^2}} f_P  .
\label{eqn:HAIR2}
\ee
The quantity $\mu^2_0$ is the squared mass of the
``quenched approximation eta-prime'' in the chiral limit.
(The factor  $1/N_f$ converts the single-flavor graph into the expectation
of the eta-prime mass in $N_f$-flavor QCD,
 since each closed loop has a multiplicity of $N_f$,
and the wave function (vertex) is scaled by a factor of $1/\sqrt{N_f}$.)
In full QCD the correlator which gives the mass of the isosinglet
meson is the difference $C(t)-N_f H_{full}(t)$, and $H(t)$ is supposed to
represent  the first
term in a geometric series, which sums up to
\bee
C(q) - N_f H_{full}(q) =
C(q) - N_f H(q) + \dots = f_P {1 \over {q^2+m_\pi^2+ \mu_0^2}}f_P,
\ee
shifting the squared mass of the meson from $m_\pi^2$ to $m_\pi^2 + \mu_0^2$.
In these expressions,
 $f_P = \langle 0 |\bar\psi \gamma_5 \psi  | \pi \rangle
= m_\pi^2f_\pi/(2m_q)$ from the PCAC relation.
Computing the susceptibility directly from
Eq. (\ref{eqn:HAIR2}) gives
\bee 
{1 \over V} \sum_{x,y} H(x,y) =  {f_P^2 \over m_\pi^4} {\mu_0^2 \over N_f}
 = {{\mu_0^2 f_\pi^2} \over {4 N_f m_q^2}} .
\label{eqn:HAIR3}
\ee
Equating Eqs. (\ref{eqn:HAIR1}) 
and (\ref{eqn:HAIR3}), we obtain the Witten-Veneziano
relation $\mu_0^2=4N_f\chi/f_\pi^2$, where $\chi$ is the zero mode
susceptibility.

This simple derivation of the Witten-Veneziano relation from the overlap
action uses only the quenched approximation, without any reference to
the large $N_c$ limit. The crucial ingredient is the fact that the
hairpin graph, in the quenched approximation, takes the form of
 Eq. (\ref{eqn:HAIR2}).
Conventional derivations
of the Witten-Veneziano relation in unquenched QCD proceed rather differently:
One begins by considering a correlator of the local topological charge density,
\bee
U(k) = \int d^4 x \exp(ikx) \langle
( {g^2\over{16\pi^2 N_c}}  F \tilde F(x) )
( {g^2\over{16\pi^2 N_c}}F \tilde F(0))
\rangle
\label{eq:firsta}
\ee
which is assumed to be dominated by an eta-prime resonance plus
other massive states
\bee
U(k)  = { C_{\eta'}^2\over{k^2+m_{\eta'}^2}} + \dots
\label{eq:secnda}
\ee
with $C_{\eta'} = \langle 0 | (g^2/(16\pi^2 N_c) F \tilde F | \eta' \rangle$.
Using the anomaly equation
 $ 2N_f (g^2/(16\pi^2 N_c) F \tilde F = \partial_\mu J_5^\mu$
allows one 
to replace the gluonic matrix element by the  matrix element of the
divergence of the quark
current. 
Computing the susceptibility by taking $k$ to zero
and setting the eta-prime and pion decay constants equal, so
$\langle 0 | \partial_\mu J_5^\mu |\eta'\rangle = 
\sqrt{N_f} m_{\eta'}^2 f_\pi $,
 gives the Witten-Veneziano relation. The difference
is that here the eta-prime really is a propagating particle,
whereas in the quenched approximation the mass is just the value of
the two-boson coupling. Note also that in the derivation we have just given,
the topological susceptibility is replaced by the squared
 fluctuation in  the number of fermionic
zero modes. The $k\rightarrow 0$ limit of $U(k)$ is a problematic one
due to the contribution of contact terms \cite{ref:FFDUAL}.

A third  kind of derivation of the Witten-Veneziano relation
has recently been given by
Giusti, Rossi, Testa and Veneziano \cite{ref:Giusti}.
Building on derivations of the Ward identity for the
flavor singlet axial vector current by
Hasenfratz, Laliena and Niedermayer\cite{ref:HLN}
and L\"uscher \cite{ref:Lusc},
they show that the Witten-Veneziano relation for overlap fermions involves
the zero mode susceptibility, not the topological susceptibility.
However, they assume that the axial current correlator is saturated by 
a propagating eta-prime resonance, keeping Eq. (\ref{eq:secnda}).
That does not occur in quenched approximation, where the flavor singlet
channel propagator is given by Eq.
(\ref{eqn:HAIR2}).

In full QCD the hairpin correlator is a single
particle propagator, but the correlator which gives the eta-prime
mass is given by $C(q) - N_f H_{full}(q)$. $H_{full}(q)$ still couples
to zero modes, but in order for the difference to couple only to the eta
prime, the hairpin must have a piece which cancels the connected
correlator $C(q)$, the susceptibility of which is order $1/m_\pi^2$.
Then $H_{full}(0) = 1/m_\pi^2 +\dots = \chi/m_q^2$
 and one  finds \cite{ref:FSU98,Chandrasekharan:1998wg}
the expected result \cite{ref:LS} that $\chi \simeq m_q$.
The eta-prime mass is only connected to the zero-mode
susceptibility through
terms which are higher order in the quark mass.

We will compute the zero mode susceptibility both directly and
 via a calculation of the hairpin graph, and compare the results.
In practice, in common with all lattice calculations of matrix
elements, we will extract $\mu_0^2$ from correlators using extended sources
and sinks. We infer that the mass relation persists, regardless of the
choice of source and sink, by assuming that the hairpin graph is
a two-point vertex of pseudoscalar mesons, as required by
quenched chiral perturbation theory.

We will see that  Eq. (\ref{eqn:HAIR1})  produces a measurement of $\mu_0^2$
and $\chi$
with a somewhat larger statistical uncertainty
 than one would get from
a direct measurement (counting zero modes) of $\langle Q^2 \rangle$.
This happens despite the fact that in the fit to the hairpin,
 one is using information from many points of the
lattice, while counting zero modes gives just one number per lattice.
The signal from the many points just shows very strong correlations
 from point to point, and the actual hairpin graph is computed at nonzero
quark mass and must be extrapolated to the chiral limit.

\section{The Lattice Calculation}
The overlap action used in these studies\cite{ref:TOM_OVER}
 is built from an action with nearest and
next-nearest neighbor couplings,
and  APE-blocked links\cite{ref:APEblock}.
Eigenmodes of the massless
 overlap Dirac operator $D(0)$ are constructed from eigenmodes
of the Hermitian Dirac operator $H(0)=\gamma_5 D(0)$,
 using an adaptation of
a Conjugate Gradient algorithm of Bunk et. al.
and  Kalkreuter and Simma\cite{ref:eigen}.
These eigenmodes are used to precondition
the calculation of the quark propagator and to construct quark propagators
truncated to some number of low lying eigenmodes.

The data set is generated in the quenched approximation
 using the Wilson gauge action at a coupling
 $\beta=5.9$. The nominal lattice spacing is $a=0.13$ fm
from the measured rho mass (a value we prefer; see below) or
 inferred to be 0.11 fm from the Sommer parameter
using the interpolation formula of Ref. \cite{ref:precis}.
We worked with lattices with $12^3 \times 24$ sites.
Quark masses in lattice units are $am_q=0.02$, 0.04, and 0.06,
corresponding to pseudoscalar-to-vector meson mass ratios
of $m_{PS}/m_V \simeq0.5$, 0.61 and 0.67.
The fermions have periodic boundary conditions in the
 spatial directions and anti-periodic temporal boundary conditions.
We calculated the twenty  smallest eigenvalue modes of $H^2(0)$ in the chiral
sector of the  minimum eigenvalue,
 and reconstructed the degenerate opposite chirality eigenstate
of $H^2(0)$ for each nonzero eigenvalue mode.  These modes are
then recoupled into eigenmodes of $D(0)$.
Their eigenvalues have imaginary parts ranging up to $0.3/a$-$0.35/a$,
or about 500 MeV \cite{ref:dh00}.

All correlators we measured will include a smearing function which
averages  the source of the propagator  over a localized spatial volume,
in order that the meson source resembles an actual hadron wave function.
The hairpin correlator that we measure is
\bee
H_\Gamma(t) = {1 \over T}\sum_{t_1}
   \langle \sum_{x} \sum_{y} \Tr L_\Gamma(x,t+t_1) \Tr L_\Gamma(y,t_1) 
 \rangle 
\ee
where the single fermion loop is
\bee
L_\Gamma(x,t) = \sum_{x_1,x_2} \Phi(x_1-x)\Phi(x_2-x)
\langle \bar \psi(x_1,t) \Gamma \psi(x_2,t) \rangle .
\ee
$\Gamma$ is a product of Dirac matrices: $\gamma_5$ 
 and $\gamma_0 \gamma_5$ are studied
for the eta prime.

The connected correlator uses the same (separable) weighting function
\bee
C_\Gamma(t)= \langle \sum_{x}\sum_{x_1,x_2}  \Phi(x_1-x)\Phi(x_2-x)
\bar\psi(x_1,t)\Gamma\psi(x_2,t) 
\sum_{y_1,y_2}\Phi(y_1)\Phi(y_2)
\bar\psi(y_1,0) \Gamma\psi(y_2,0) \rangle .
\ee
We take the weighting function to be a Gaussian,
$\Phi(x)=\exp(-|x|^2/r_0^2)$, of width $r_0=3a$.

In coordinate space we fit 
\bee
C(t) = {Z \over {2m}}(\exp(-mt)+\exp(-m(T-t))
\label{eq:ct}
\ee
and, since
\bee
H(q) = - {\mu_0^2 \over N_f} {{\partial }\over{\partial m^2}}C(q),
\ee
the coordinate space hairpin is
\bee
H(t)= {{Z }\over{4m^3}}{\mu_0^2 \over N_f} 
((1+mt)\exp(-mt)+(1+m(T-t))\exp(-m(T-t))).
\label{eq:ht}
\ee

We break up the computation of the single loop into two terms.
The first term includes the contribution of the low eigenmodes of $D$,
 and is done exactly (for those modes):
Chiral eigenmodes of $D(0)^\dagger D(0)$ are labeled as $\phi_{j\pm}(x,t)$,
the plus or minus sign corresponding to the chirality.
They have associated eigenvalues $\lambda_j$. These modes are recoupled
into eigenmodes of $\hat D^{-1}(m)$. The only mode mixing occurs between
degenerate eigenmodes of $D(0)^\dagger D(0)$. The
 single fermion loop becomes
\bee
L_1(x,t) = \sum_j^N \left[ \Tr \ \alpha_{j+} \Psi^\dagger_{j+}(x,t) \Gamma
\Psi_{j+}(x,t)
 + \Tr \ \alpha_{j-} \Psi^\dagger_{j-}(x,t) \Gamma \Psi_{j-}(x,t) \right ] ,
\label{eq:L_1}
\ee
where
$\Psi_{j\pm}(x,t) = \sum_{x_1} \Phi(x-x_1)\phi_{j\pm}(x_1,t)$ is the
convolution of the $j$th eigenmode
 with the smearing function $\Phi$.
Here, defining $\mu=m/(2x_0)$, $\epsilon_j= \lambda_j/(2x_0)$, the
$\alpha_{j\pm}$, the eigenvalues of $\hat D^{-1}(m)$, are
\bee
\alpha_{j\pm}={1\over{2x_0}}
{{\mu(1-\epsilon_j^2) \pm i \epsilon_j \sqrt{1-\epsilon_j^2} } \over
 {\epsilon_j^2 + \mu^2(1-\epsilon_j^2)}} .
\ee
For zero modes only one term in the brackets in Eq. (\ref{eq:L_1}) exists.

The second part of the loop is computed using a stochastic estimator.
We cast a vector of Gaussian random numbers
 $|r(x,t) \rangle$ on every site of the lattice, projected the
modes used in $L_1$ from it (so $|r_1 \rangle = |r\rangle -
\sum_j |j \rangle \langle j|r\rangle$), convoluted $|r_1 \rangle $ with
the smearing function, and used this vector as the source for
a propagator. A final convolution of the propagator with the source
vector produced a noisy estimator for the difference $L(x,t)-L_1(x,t)$.
We averaged over twelve random sources per lattice. Each source was broken into
two opposite chirality pieces, so a total of 24 inversions, restricted to
a single chirality sector each, were done per configuration.

In retrospect, our simple stochastic estimator did not produce a useful signal.
We first generated a 40-lattice data set, on which we found the
complete hairpin correlator, from both low modes and stochastic estimator
of high modes.
 For the pseudoscalar hairpin, all of the signal
is contained in the low eigenmodes, which we captured exactly. Even so,
we did not have a good enough signal from 40 lattices to analyze the hairpin
graph, so we generated an additional 40 lattices. On these lattices we 
did an ordinary spectroscopy calculation for the connected diagram
and computed the hairpin from the low modes. For our version of the overlap,
the cost of finding the eigenmodes in addition to constructing ordinary
propagators is neglegible because  the eigenmodes can be used to precondition 
the  Conjugate Gradient calculation of propagators.
At 24 inversions per lattice, a stochastic evaluation of the higher modes'
contribution is about twice as costly as an ordinary propagator calculation.

We did not experiment with more sophisticated noise reduction methods
\cite{ref:NOISE}.

\section{Results}
\begin{figure}[thb]
\begin{center}
\epsfxsize=0.8 \hsize
\epsffile{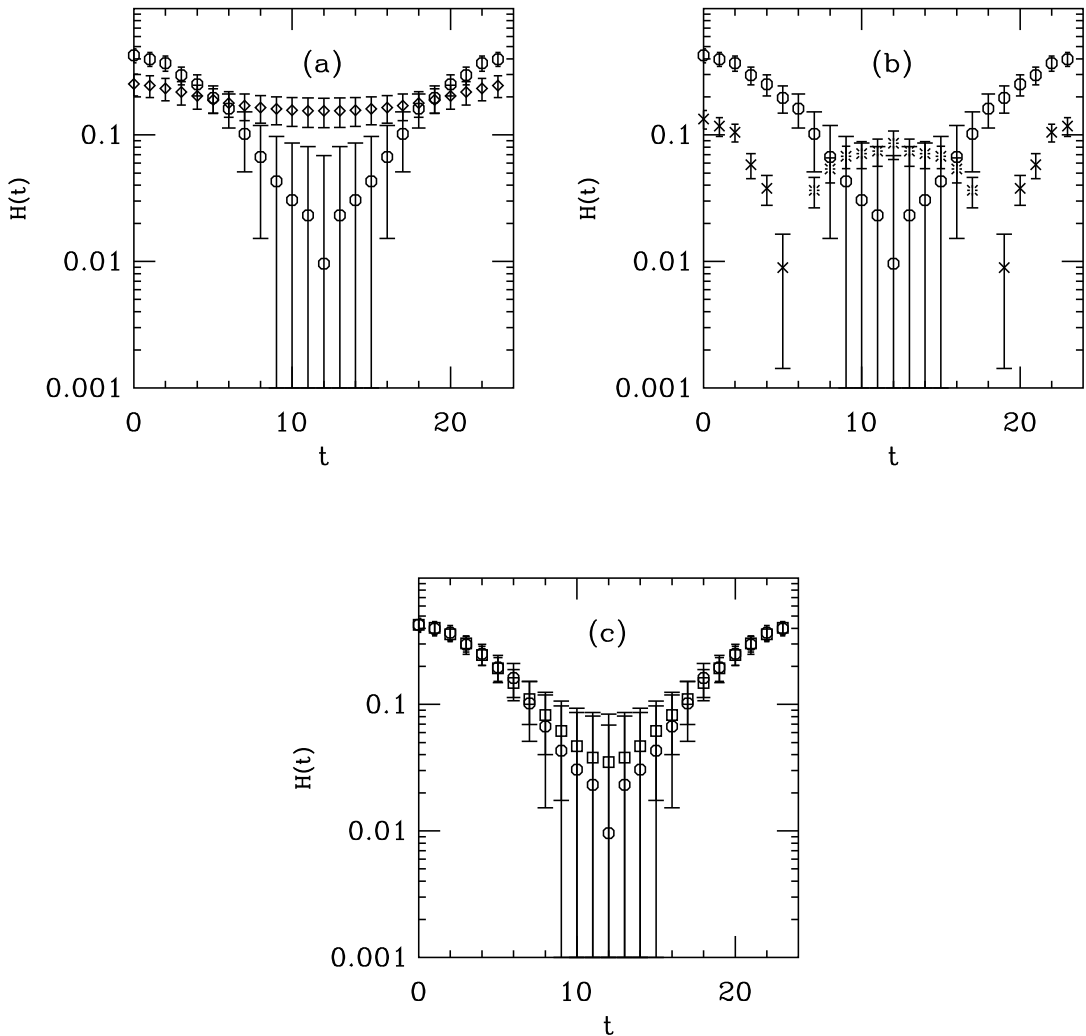}
\end{center}
\caption{
Contribution to the pseudoscalar hairpin for the $am_q=0.04$ data set
 from various subsets of eigenmodes:
In all figures the full propagator is shown by octagons. 
In panel (a), the zero mode contribution is
given by diamonds.  In (b) 
the contribution of all nonzero modes is shown by
crosses (where the contribution is positive) and bursts (where it is negative).
 In (c)
the contribution of the lowest 20 modes of $D^\dagger D$ is shown
by squares.
}
\label{fig:propsPS}
\end{figure}

It is an interesting exercise to ask how the contributions of different sets
of eigenmodes contribute to the pseudoscalar
hairpin correlator.
We find that the low modes completely saturate the hairpin graph at all
 mass values studied. Because the plots are rather cluttered, we show 
results for only one mass value, $am_q=0.04$,
in Fig. \ref{fig:propsPS}:
The full correlator, including the low modes treated exactly and the high
modes computed using the noisy estimator, is shown by octagons.
Completely overlapping these points is the correlator built of the lowest
20 eigenmodes of $D^\dagger D$.

The zero mode piece is shown by diamonds. 
 At large quark mass there is a rather strong interference
between the zero mode contribution and the nonzero modes, which pulls
the full propagator down. Since the low modes themselves saturate
the correlator, this interference must be dominated by them.

Finally, the contribution of the nonzero modes by themselves is
 positive at short
distance but becomes negative above a separation of
$t \ge 7$ lattice spacings. We have displayed this by changing the plotting
symbol for this contribution from crosses, when the signal is positive,
to bursts, where it becomes negative.

It is easy to understand why nonzero modes contribute negatively at large
separation.  
The $j$th pair of nonchiral modes contribute a term to the hairpin of
\bee
\hat H(t)= {1\over T}\sum_{t'} \sum_{j,j'} \sum_{x,x'}\alpha_j \alpha_{j'}
 \langle \Omega_j(x,t+t') \Omega_{j'}(x',t') \rangle
\ee
where (see Eq. (\ref{eq:L_1}))
\bee
\Omega_j(x,t) = \sum_{x_1,x_2} \Phi(x_1-x)\Phi(x_2-x)
\phi_j(x_1,t)^\dagger \gamma_5 \phi_j(x_2,t)
\ee
is a local smearing of the chiral density of the $j$th mode.
The nonzero modes have zero overall chirality and so at large distance
the correlation function of the chiral density must become negative.
This can be seen even on a single configuration.
This behavior has been observed in the eigenmode studies of
 Ref. \cite{ref:dh00}. It may be more than a coincidence that these authors
 saw the chirality autocorrelator (for the same data set as we studied here)
becoming negative around a distance of 6 or 7 lattice spacings.

\begin{figure}[thb]
\begin{center}
\epsfxsize=0.5 \hsize
\epsffile{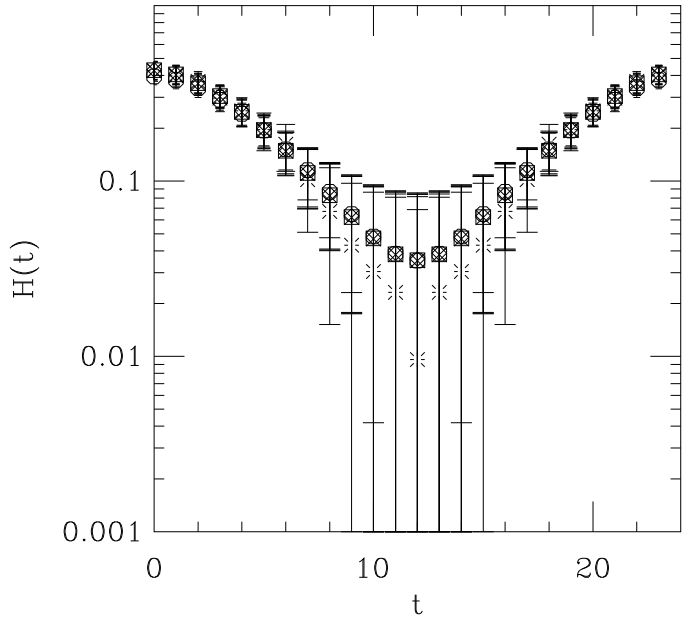}
\end{center}
\caption{
Saturation of the pseudoscalar hairpin graph at $am_q=0.04$
 by 5, 10, 15, and 20 eigenmodes
of $H(0)^2$ (with symbols octagons, diamonds, squares, and crosses).
Bursts show the full correlator.
}
\label{fig:low}
\end{figure}

Could we have used fewer modes?
Fig. \ref{fig:low} shows the contributions to the hairpin correlator of
the lowest 5, 
10, 15, and 20 modes of $H(0)^2$, for the $am_q=0.04$ data set.
The pictures for the other masses are virtually identical. We could have
 kept only five modes of $H(0)^2$ and almost completely
saturated the hairpin correlator, in our particular simulation volume.

\begin{figure}[thb]
\begin{center}
\epsfxsize=0.5 \hsize
\epsffile{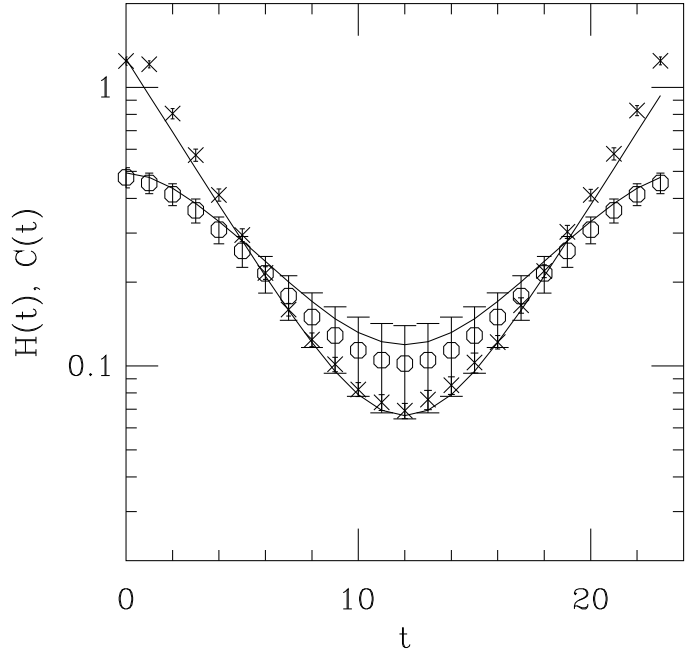}
\end{center}
\caption{
Connected correlator (crosses) and hairpin (octagons), for the
lowest 20 eigenmode hairpin, at bare quark mass $am_q=0.04$, showing
the result of a correlated fit
 to the two correlators over the range $t=6$ to 18 (folded to $6-12$).
}
\label{fig:fitPS04}
\end{figure}

\begin{figure}[thb]
\begin{center}
\epsfxsize=0.5 \hsize
\epsffile{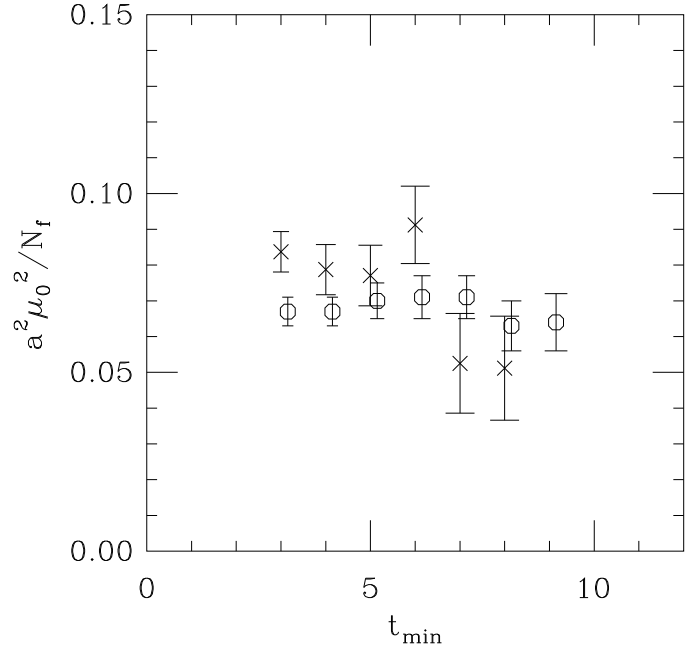}
\end{center}
\caption{
Extracted pseudoscalar 
hairpin coupling $\mu_0^2/N_f$ from fits from distance $t$ to 12
at $am_q=0.04$.
Octagons show the results of fits to the hairpin correlator when the
 quark propagator
is  truncated to the
lowest 20 eigenmodes (80 lattice data set)
of $D^\dagger D$, and crosses show fits using the full 
 propagators (40 lattice data set)
 in the hairpins.
}
\label{fig:mrangePS}
\end{figure}

We have performed correlated fits to the connected and hairpin pseudoscalar
 graphs to Eqs. (\ref{eq:ct}) and (\ref{eq:ht})
and extracted lattice predictions for $\mu_0^2/N_f$.
An example of
a fit, to the $am_q=0.04$ low-eigenmode hairpin, is shown in Fig.
 \ref{fig:fitPS04}.
The error bars on the points are the naive ones and do not show the
considerable correlation between the data from different time slices.
A set of fits, 
for a time slice range of $t_{min}=3$  out to $t_{max}=12$, is shown
in Fig. \ref{fig:mrangePS}. We have separated fits to using hairpin correlators
with truncated quark propagators and fits using the full quark
 propagator in the hairpin. All of the fits to the low-mode hairpin have
good confidence levels for $t_{min} \ge 5-6$.
Pion masses are also a byproduct of the fit. Their values are stable for
$t_{min} \ge 3-4$ and equal within uncertainties to the results of fits to
pion propagators in isolation.

\begin{figure}[thb]
\begin{center}
\epsfxsize=0.5 \hsize
\epsffile{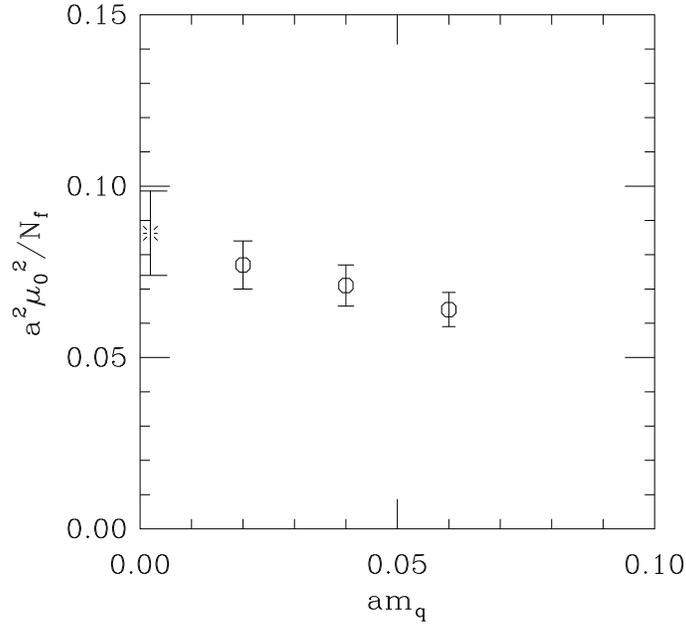}
\end{center}
\caption{
Pseudoscalar 
hairpin coupling $\mu_0^2/N_f$ in lattice units as a function of quark mass.
The jackknife-extrapolated zero-quark-mass value is shown as the burst.
}
\label{fig:mrangeeta}
\end{figure}

The ``full'' signal is much noisier than the low mode truncation
(even more than one would expect, knowing that the low mode data
 set is twice as big), and,
since both are equal within uncertainties, we quote the latter.
Our chosen best fits, typically over a range 6-12, are shown in
Table \ref{tab:results} and plotted in Fig. \ref{fig:mrangeeta}.
We have also extrapolated our results for $\mu_0^2/N_f$ to zero quark mass,
assuming that $\mu_0^2/N_f$ is a linear function of the quark mass,
using a single elimination jackknife.

Taking the lattice spacing from the rho mass, {\it i.e.} $a=0.13$ fm,
and setting $N_f=3$ we predict $\mu_0 = 770(54)$ MeV $\times
(a^{-1}/1520$ MeV). Had we used the Sommer parameter to set the scale,
{\it i.e.} $a=0.11$ fm or $a^{-1} = 1770$ MeV, the mass would have come
out about 16\% higher. This difference is, of course, a reflection of the
inherent scale uncertainty in quenched QCD.
Evaluating the left-hand side of Eq. (\ref{eqn:WV}) with
observed particle masses gives $\mu_0 = 860$ MeV.

\begin{figure}[thb]
\begin{center}
\epsfxsize=0.5 \hsize
\epsffile{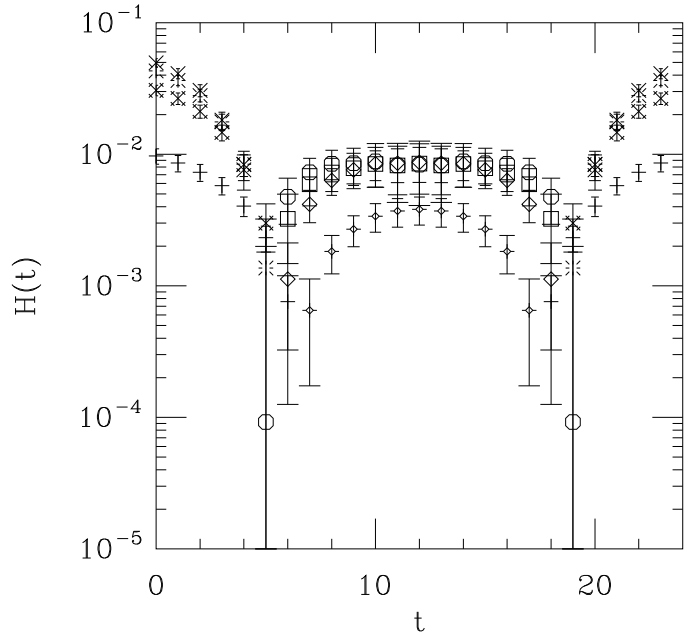}
\end{center}
\caption{
Contribution to the  $am_q=0.04$ $\gamma_0 \gamma_5$
 hairpin from low modes of $D^\dagger D$: 5 modes (small diamond with cross
when positive, plus sign when negative); 
10 modes (diamond and fancy cross), 15 modes (square and burst) and 20
modes (octagon and cross).
}
\label{fig:propsPS004}
\end{figure}

If the ``standard picture'' of the hairpin graph
 makes sense, we should be able to measure $\mu_0^2$ using
any interpolating field at the source and sink points. In particular, the
axial vector ($\gamma_0 \gamma_5 - \gamma_0 \gamma_5 $) 
hairpin correlator is an interesting operator to
study. No zero modes contribute to its hairpin correlator, and yet
the pseudoscalar and
axial vector hairpins should both give the same result for $\mu_0^2$ --
the only difference in the two channels is that the coupling of
 the external current source to the pseudoscalar channel is different.

We attempted to measure $\mu_0^2$ from
this correlator, without success. We raise
the point as a potentially interesting exercise if future high-statistics
data become available.

At short distances this correlator is negative. At larger
distances it swings positive, before
disappearing into noise at large separation. (See Fig. \ref{fig:propsPS004}
for the $am_q=0.04$ data set.)
The node in the correlator makes it impossible to fit the data to
the simple form of the hairpin we have used above.
Notice from Fig. \ref{fig:propsPS004} that the hairpin correlator
saturates more slowly with number of eigenmodes, than the pseudoscalar 
correlator does: this, plus the nodes in the signal, makes the analysis
impossible to do.

\begin{figure}[thb]
\begin{center}
\epsfxsize=0.5 \hsize
\epsffile{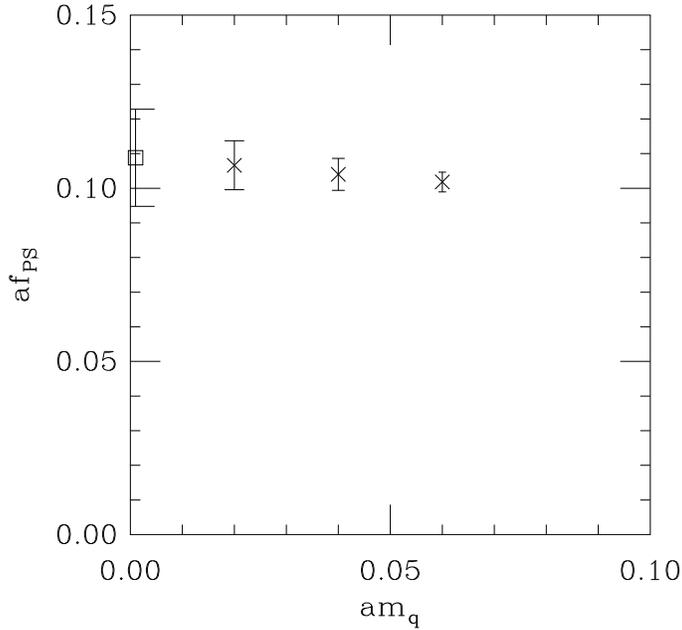}
\end{center}
\caption{
Pseudoscalar decay constant (in lattice units) from this overlap action.
Crosses are our data and a jackknife linear extrapolation to
 $m_q=0$ is shown by
the square.}
\label{fig:fpi}
\end{figure}

Now we return to the pseudoscalar correlator.
To make contact with other chiral observables we need a lattice determination
of $f_\pi$. We do this by measuring the pseudoscalar density $P$ (as one
vertex of a two-point function) and extracting $f_\pi$ from
$m_\pi^2 f_\pi = 2 m_q \langle 0|P |\pi \rangle$.
 Because the overlap action is chiral, the renormalization factors for
the quark mass and the pseudoscalar density cancel.
The pseudoscalar density is taken to be the naive operator
 $P(x) = \bar \psi \gamma_5 \psi$.
Note that because we use the ``subtracted'' propagator $\hat D^{-1}$,
Eq, (\ref{SUBPROP}),
our measurement is equivalent to the use of the ``order $a^2$ improved
current'' $\bar \psi (1 + D/(2x_0))\gamma_5 (1 + D/(2x_0)) \psi$.
The results of this exercise  are shown in Fig. \ref{fig:fpi} and also
given in Table \ref{tab:results}. Using again the lattice spacing from
the rho mass we find in the chiral limit, obtained by linear extrapolation
in the quark mass, $f_\pi = 163(21)$ MeV. Had we
used the Sommer parameter to set the scale we would have obtained
$f_\pi = 193(25)$ MeV. The latter number is
obviously much larger than the experimental
value of 132 MeV. We prefer the lattice spacing from the rho mass
 for phenomenology
with the overlap fermions.
A comparison of the point-to-point correlator  in the pseudoscalar channel
with instanton liquid model predictions also requires
a large lattice spacing \cite{ref:TOM3} of 0.13 fm.
This correlator is proportional to 
$\langle 0|P |\pi \rangle^2$.

\begin{table}
\begin{tabular}{|c|l|l|l|l|}
\hline
$m_0$ & $a^2 \mu_0^2/N_f$ & $a f_\pi$  & $a^4 \chi$  &  $\delta$   \\
\hline
0.06  & 0.064(5) & 0.102(2)& $1.67(21)\times 10^{-4}$ & 0.077(8)\\
0.04  &0.071(6) & 0.104(4) & $1.95(34)\times 10^{-4}$ & 0.083(17) \\
0.02  & 0.077(7) &0.106(7) & $2.24(40)\times 10^{-4}$  & 0.088(20)\\
\hline
   0 & 0.086(12) & 0.108(14) & $2.52(58)\times 10^{-4}$ & 0.093(28) \\
\hline
\end{tabular}
\caption{Table of best-fit lattice parameters $a^2 \mu_0^2/N_f$, $a f_\pi$,
 $a^4 \chi$, and $\delta$
as a function of quark mass and after jackknife extrapolations to zero mass.}
\label{tab:results}
\end{table}

\begin{figure}[thb]
\begin{center}
\epsfxsize=0.5 \hsize
\epsffile{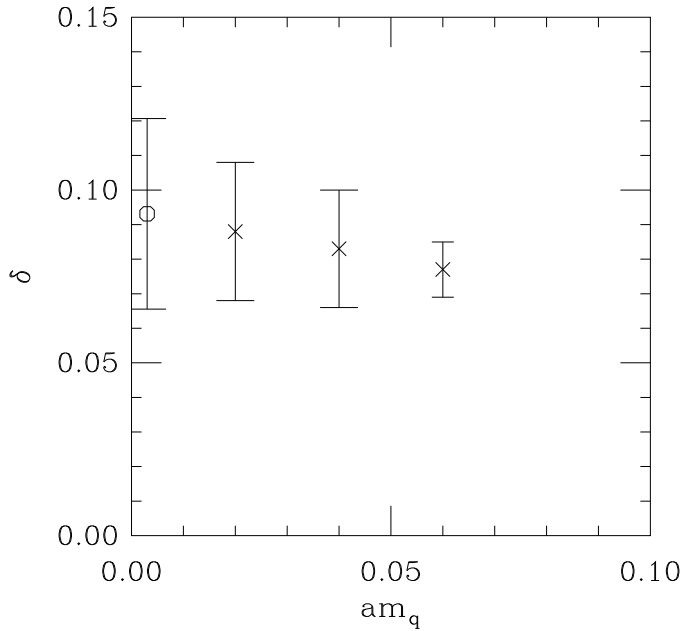}
\end{center}
\caption{
The quenched chiral perturbation theory parameter 
$\delta=\mu_0^2/(8 \pi^2 N_f f_\pi^2)$ evaluated for $N_f=3$.
Crosses are our data and the extrapolated value at $m_q=0$ is shown by
the octagon.}
\label{fig:delta}
\end{figure}

\begin{figure}[thb]
\begin{center}
\epsfxsize=0.5 \hsize
\epsffile{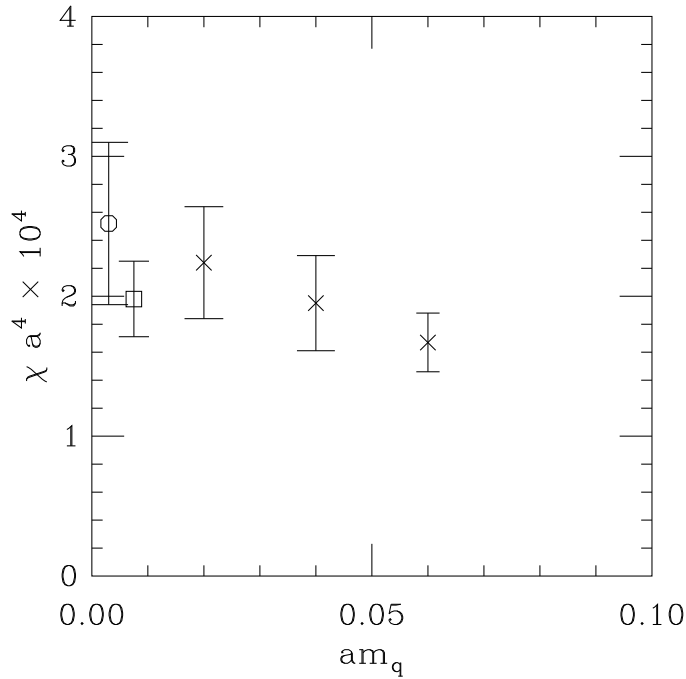}
\end{center}
\caption{
The inferred zero mode susceptibility in lattice units
from the hairpin graph 
 $ \chi = \mu_0^2 f_\pi^2/(4 N_f)$ evaluated for $N_f=3$.
Crosses are our data and the extrapolated value at $m_q=0$ is shown by
the octagon. The susceptibility measured directly from zero modes is
shown with the square.
}
\label{fig:chi}
\end{figure}

The two parameters which are related to $\mu_0^2$ are the zero mode
susceptibility 
 $ \chi = \mu_0^2 f_\pi^2/(4 N_f)$ and the parameter which
 characterizes the strength of the hairpin contribution in quenched chiral
perturbation theory \cite{ref:chiralpt},
 $\delta=\mu_0^2/(8\pi^2  N_f f_\pi^2)$.
We compute each of these parameters for each value of quark mass using
a combination of single-elimination jackknife with correlated fits
to pairs of two-point functions, and extrapolate the results linearly
in the quark mass to $m_q=0$. The resulting values are shown in Table 
\ref{tab:results} and in Figs. \ref{fig:delta} and \ref{fig:chi}.

We can also perform a direct measurement of the topological susceptibility
by counting zero modes: the topological charge $Q$ on each configuration is
just defined as the difference in number of opposite chirality zero modes
in the configuration.
 For our 80 lattices, we find
$\langle Q \rangle = 0.31 \pm 0.32 $.
We can (formally) eliminate the $\langle Q \rangle^2$ term from the
average by combining our data with a parity-reversed copy of every lattice.
(This is equivalent to the usual replacement of a propagator by its real
part in conventional spectroscopy or matrix element calculations.)
The zero modes give us the  result
$\langle Q^2 \rangle = 8.24 \pm 1.11 $, and
\bee
a^4 \chi = {{\langle Q^2 \rangle }\over V} =
1.98(27)\times 10^{-4}   .
\ee
This result agrees 
 with the calculation of $\chi$ from $\mu_0^2$.
The quality of our signal is similar to what is seen in
calculations of $\chi$ using pure gauge observables
(lattice analogs of the topological charge density)
with similar statistics. 
Converting this lattice number to a continuum result using the Sommer
scale, as is conventional for the topological susceptibility,
 and taking the fourth root yields $\chi^{1/4} = 213(7)$ MeV
from the direct counting of zero modes and
$\chi^{1/4} = 226(12)$ MeV from the hairpin fit.
Pure gauge studies produce similar numbers to what we see \cite{ref:TEPREV}.

Obviously, with the scale set by the rho mass, the result would be
about 16 per cent lower, 180(6) and 191(11) MeV.

Our measurement of $\delta\simeq 0.093$
is  a bit higher than recent determinations made using nonchiral
actions\cite{Bardeen:2000cz,Aoki:2000yr,Bernard:2001av}, which measure
 $\delta\simeq 0.065$ (but somewhat smaller than what one would infer
from the W-V relation with physical masses, Eq. (\ref{eqn:WV})).
The first calculation is done using clover fermions
 at $\beta=5.7$. Clover fermions have real eigenmodes which 
 make a nonvanishing contribution to $\Tr \gamma_5 D^{-1}$.
The overlap action counts each zero mode with unit weight.
It is our experience that as the value of the real eigenvalue
of a clover mode moves away from zero, the associated chirality
also decreases, and so it is not surprising that our result is larger.
The other two calculations determine $\delta$ from the small quark mass
behavior of the pion mass, {\it i.e.} from observing quenched chiral
logarithms, and can easily have quite different systematic uncertainties
than our measurement.

We do not see any effects of quenched chiral logarithms in any
observables. 
A fit to the pion mass in
 various channels (pseudoscalar, axial vector, difference of
pseudoscalar and scalar) to the form suggested in Ref. \cite{Bernard:2001av},
$m_\pi^2 = Cm_q(1 - \delta \log (C m_q/\Lambda^2))$, chooses $\delta$'s
in the range 0.0-0.2, but with uncertainties from any fit also of the same
order.
The main effect of quenching we have seen is the presence of zero modes
in the pseudoscalar and scalar
channels; in contrast to the hairpin case, these are
finite volume artifacts.

\begin{figure}[thb]
\begin{center}
\epsfxsize=0.5 \hsize
\epsffile{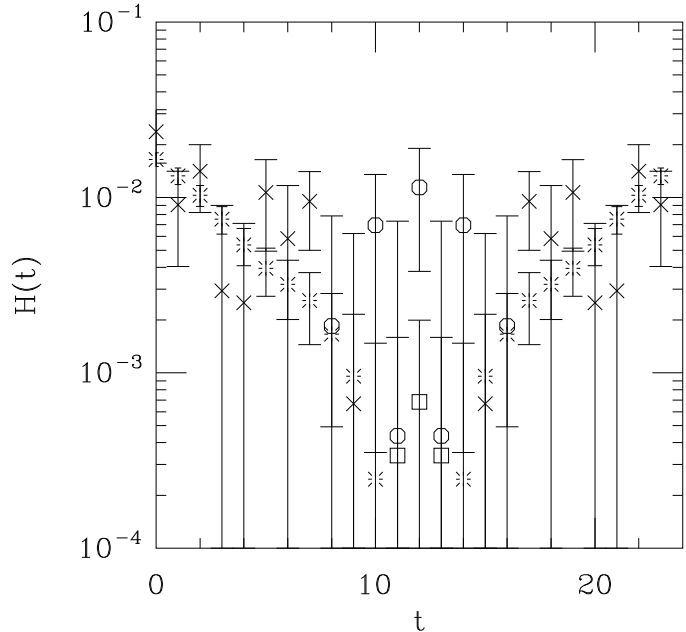}
\end{center}
\caption{
Contribution to the  $am_q=0.04$ $\gamma_i$
 hairpin from the lowest 20  modes (square
when positive, burst when negative) as compared to the full correlator
(octagon when positive, cross when negative).
}
\label{fig:propsV04}
\end{figure}

We do not have any useful signals from other channels. The scalar channel
is dominated by zero and low eigenmodes, but the major structure in the channel
is just the constant chiral condensate.
In the vector channel the low-eigenmode correlator even has the
wrong sign--it is negative over most of its range
 (compare Fig. \ref{fig:propsV04}).
The absence of a signal could be consistent with a naive expectation from
Zweig's rule, namely that the size of the hairpin in the
 vector channel is small.
The low eigenmodes which we include exactly do not seem to make much of
a contribution to disconnected diagrams in the vector channel, either.
This feature is expected in instanton liquid models \cite{Schafer:1996wv}.

We would feel much more comfortable making these statements if we had a
 real signal \cite{Isgur:2000ts}.

\section{Conclusions}
For  overlap actions, and in the quenched approximation, 
 the Witten-Veneziano formula is an exact relation between the eta-prime mass
inferred from the hairpin graph and 
the fermionic zero-mode susceptibility.
However, the physics of the Witten-Veneziano relation in the overlap
is rather different than in ``standard derivations:'' the
eta-prime is not a real particle; the mass we measure
is the size of a quenched-artifact coupling between two flavor-singlet
Goldstone bosons. For quenched QCD,
 the gauge configurations which ``fill in the white space''
in the hairpin diagram are the ones which produce zero modes.
We have not addressed the question of what choice of contact term
is needed to equate a particular definition of the topological susceptibility
to the zero mode susceptibility.

The numbers we have found for the inferred eta-prime mass and
the zero mode susceptibility are not all that different from previous
results using nonchiral fermion actions. We believe that the theoretical
underpinning of our calculation done with a chiral action
is more reliable than any calculation done with a nonchiral action.
We are well aware that our calculation is performed at only one value of
the lattice spacing, and that simulations at several lattice spacings
are necessary for an honest extrapolation to a continuum value.

The strong coupling of low eigenmodes to the hairpin amplitude
allowed us to perform the numerical simulation.
This connection between the zero mode susceptibility and the
coupling strength is not automatic for a nonchiral fermion action.
Nevertheless, we expect that small eigenmodes of the Dirac operator will
make a large contribution to the pseudoscalar hairpin correlator.
We suggest  that future
studies of the hairpin graph, even done using nonchiral actions, can
reduce the measurement noise by 
 first finding all the zero or near-zero eigenmodes of the
 Dirac operator so that their contribution can be included
 exactly  \cite{Schilling:2001xd}.

\section*{Acknowledgements}
This work was supported by the
U.~S. Department of Energy with grants DE-FG02-97ER41022 (UMH) and
DE-FG03-95ER40894 (TD). The computations were carried out on Linux clusters
at CU and FSU.


\end{document}